# A USER PROFILE BASED ACCESS CONTROL MODEL AND ARCHITECTURE

Meriem Zerkouk[1], Abdallah Mhamed[2] and Belhadri Messabih[1]

[1]University of Sciences and Technology Oran, Algeria.
`{zerkouk.meriem, bmessabih}@ gmail.com`

[2] Institut Mines-Télécom / Télécom SudParis - CNRS Samovar UMR 5157 Evry - France.
`abdallah.mhamed@it-sudparis.eu`

*ABSTRACT*

*Personalization and adaptation to the user profile capability are the hottest issues to ensure ambient assisted living and context awareness in nowadays environments. With the growing healthcare and wellbeing context aware applications, modeling security policies becomes an important issue in the design of future access control models. This requires rich semantics using ontology modeling for the management of services provided to dependant people. However, current access control models remain unsuitable due to lack of personalization, adaptability and smartness to the handicap situation.*

*In this paper, we propose a novel adaptable access control model and its related architecture in which the security policy is based on the handicap situation analyzed from the monitoring of user's behavior in order to grant a service using any assistive device within intelligent environment. The design of our model is an ontology-learning and evolving security policy for predicting the future actions of dependent people. This is reached by reasoning about historical data, contextual data and user behavior according to the access rules that are used in the inference engine to provide the right service according to the user's needs.*

## 1. INTRODUCTION

Development and innovation of new assistive technologies and services are in continuous progress to provide help for the daily life activities. Ambient assisted living and context awareness allow dependent people to perform their required services in their living spaces for various applications (healthcare, home and transport). It meets to the needs of disabled and older people by making their life easier to overcome barriers for studying, working and living activities. Our aim is to address in our proposed approach the adaptability and personalization aspect in ambient assisted environment. The adaptability must rely on services which become more and more personalized with respect to the assistive devices technology and the user's needs. Service provision is handled by the development of context aware based access control models for pervasive systems. The security policy must be adaptive to the potential changes which may occur over the space and the time. This is accomplished by extending the most popular Role based access control model (RBAC) [1] and Organization Based Access Control model (OrBAC) [8]. The permissions are assigned according to the validity of context, which is a key element in the design of access control models taking into account the situation evolution in the environment. According to our literature review the current access control policies do not take into account the user impairments nor the behavior of users. Furthermore the authentication means (username/password, RFID, context and other) used to identify the user and to control their access to services can be forged or replayed because they are not time dependent.

Our work is motivated by the following challenging tasks:

1°) Providing a better identification of users based on their capabilities and behavior to ensure more suitable personalization.

2°) Using the different gathered and inferred data by the security policy specification to ensure security services.

3°) Assigning correctly the users on similar characteristics and deduce the suitable decision about the user.

In this paper, we present and describe a novel adaptive model and its related architecture. The model is based on ontology learning, enriching and evolution which support continuous learning of behavior and capability patterns. We designed an initial ontology driven knowledge base representing the access rules, context, behavior pattern, services, devices and environment. We used ontology modeling to ensure sharing, reuse, interoperability, flexibility and adaptability of the security policy. In order to implement the model, we have proposed an architecture that it is built using three layers acquisition, management and security which will be described latter in section 6.

Our approach has the following four strengths. Firstly, we identify the user correctly by analyzing his previous and present behavior. Secondly, we track the behavior to control access continuously over time. Thirdly, the model ensures security in smart environment by taking into account the different main involved entities: users, services, devices and environment. Fourthly, we provide an adaptive security policy.

The reminder of this paper is organized as follows: Section 2 analyzes the requirements of access control models and their limitations in pervasive systems. Section 3 presents our proposed approach and discusses the need of tracking and pattern recognition to build the user behavior profile. Section 4 details our proposed ontological model while section 5 gives the description of the related architecture. Section 6 presents the design process of behavior and capability security policy through a three scenarios and the last section concludes the paper.

## 2. RELATED WORK

From our literature study, RBAC seems as a standard and a reference for the design of access control model in pervasive environments. The model is based on a set of users, roles, permissions, operations, object entities, user-role and role-permission relationships [1]. The model is defined by four components: core RBAC, hierarchical RBAC, static separation of duty and dynamic separation of duty. Context is a key challenge in ubiquitous computing. Therefore, the most popular definition of context extracted from [2] is: *Context is any information that can be used to characterize the situation of an entity. An entity is a person, a place or an object that is considered as relevant to the interaction between a user and an application themselves.* Furthermore, the contextual data varies according to the context awareness environment like hospital, home, and work place. Context aware based Access control relies on context data to assign the permission to the users (roles) in the right situation which makes the model dynamic according to the change of context over the time.

Extended RBAC models [3] are based on context awareness. Their aim is to improve RBAC by assigning the right access more dynamically. The access is based on the validity of the context by adding to RBAC a single contextual data which is spatial, temporal or environmental [4, 5 and 6].

The OrBAC model is designed to overcome existing problems in extended RBAC models. OrBAC [7] adds an organizational dimension and separates between the concrete level (user, object, action) and the abstract level (roles, views, activities). It also models alternative policies (permission, prohibition and obligation, recommendation). This model incorporates different context data which can be historic, spatial, temporal or declared by user. The weakness of the model is the lack of handling the interoperability and the heterogeneity. Multi-OrBAC [8] is an extension of the OrBAC model designed for multi-organizational, its drawback lies in the fact that each organization must know the resources of the other. Poly-OrBAC [9] addresses this problem by integrating the OrBAC model to represent the internal policies of each organization and web services technology to ensure interoperability between organizations.

Privacy is a key issue in smart environment and is considered as an important feature to ensure when specifying a security policy in such pervasive systems. Privacy is considered as *"the right of the individual to be protected against intrusion into his personal life or affairs, or those of his family, by direct physical means or by publication of information"* [10]. It was used by extended RBAC based on privacy awareness in order to protect the confidentiality of users because there is a growing necessity to share the personal data between different entities in smart environment.

With the diversity of entities in the smart environment, there is another key challenge: How can we recognize and rely correctly the right person? This issue can be solved by introducing a new concept of trust. The trust level is considered as *"A particular level of the subjective probability with which an agent assesses that another agent or a group of agents will perform a particular action, both before he can monitor such action (or independently of his capacity ever to be able to monitor it) and in a context in which it affects his own action"* [11]. This key concept was included in RBAC design by proposing trust-BAC model [12] in which the role is assigned according to the trust level.

## 3. PROPOSED APPROACH

Our proposed approach aims to assist the dependent people in their living space by providing an intelligent solution to assist this category of people. The living space is equipped by following sensors to track the behavior people. Once we had collected the data from sensors, we define some rules that combine the different data to deduce a new knowledge and to classify the behaviors into behavior classes. When the user arrives, we must affect the user characterized by his contextual data to one of the defined classes then we analyze the current contextual data to identify the person (authenticated or not) then we attribute to the person (permission, interdiction, recommendation or obligation) decision. In order to implement this approach we will need an access control model and an architecture supporting them.

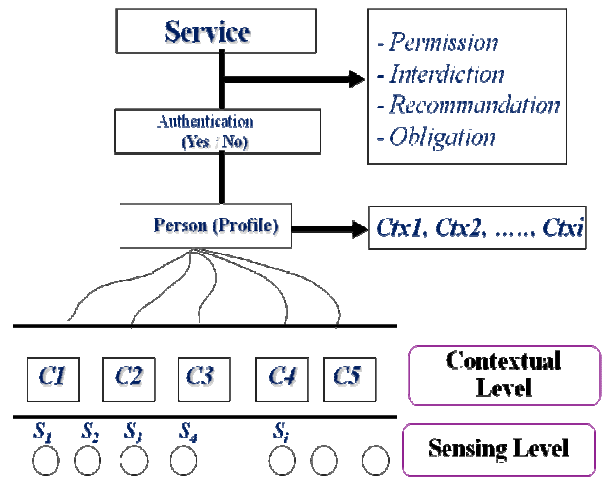

Figure 1.    Proposed approach

In order to improve the security of dependent people living in smart environment, we add the features listed above to improve the access control.

**Authentication:**  We move to the personalization for the dependent person and the identification means used are mostly badge, password and biometric data. These present credentials do not provide an appropriate personalization and good identification about user profile. Firstly, we need to identify correctly the person by checking the user capability and profile according to the behavior monitoring. Secondly, to derive the suitable authentication decision by using some defined rules.

**Authorization:** This is mean for assigning the suitable decisions (permissions, prohibition, obligation and recommendation) to the allowed persons requiring an assistive service and using an assistive device. This depends on the validity of contextual, profile, capability, behavioral data and some defined rules that permit to reason on access control. The knowledge provides a better characterization of the users living in smart home and having some particular needs.
In order to ensure the following issues, we provide and construct an intelligent security policy specification following the main four steps:

**Behavior tracking:** In this step, we collect the data from the person, the environment and the activities. The home must be equipped by different sensors in each place.
**Profile capability identification:** According to the collected data, we define some discriminate factors to distinguish between the different behavior patterns.

1- Moving_Time (): this parameter consists to analyze the time taken to move between the different rooms at home by considering the time and localization values.
2- Holding_Time (): this parameter consists to analyze the time hold to perform an activity, for this we require to analyze the time, localization and activity values.

We construct the behavior patterns according to the defined parameters which the classifiers are based on the monitored behaviors. Following this analysis, we can differentiate between the autonomic and the dependent person which this latter can be either deaf, blind or physically disabled.

 Once the profile was identified, the next task consists to authenticate the users then to analyze the contextual data for attributing to each user included in behavior classes the right decision.

**Access control policy modeling and reasoning:** It consists to represent data on standard format by using ontologies to ensure the interoperability, the sharing and the reuse of security policy. The current captured data and the inference rules stored in the database to deduce a new knowledge and to check the consistency of the ontology.

**Evolving:** It consists to learn the data provided over the time from different sources in order to update the behavior classes.

## 4. PROPOSED ACCESS CONTROL MODEL

### 4.1. Security policy modeling

In this paper, we propose a new handicap situation aware based Access Control Model which provides an adaptable access control in smart environment where the users have specific behavior, profile and capability. Therefore, the assistance is required to adapt services according to the users using assistive devices. For this, we classified users according to behavior, profile, current context situation, services, devices and environment. In order to take the real entities, we use the semantic web technologies, the ontology is described with Web language (OWL) and Semantic Web Rule language (SWRL) and the Simple Protocol and RDF (Resource Description Framework) Query Language (SPARQL) queries to ask and access to the data, these tools are used to define, represent and implement our proposed model in smart environment.

As illustrated in Figure 2 the ontology is built using five principal entities: security policy, user, service, device and environment.

Our model is defined as an ontology that expresses the security policies in such smart environment. In order to ensure an adaptive security policy, we need to take into consideration five main classes: user, device, service, environment and security policy.

The security policy class permits to ensure the authentication, the access control, the trust, the priority and the conditions. Authentication subclass its goal is to identify correctly the user using behavior subclass and authentication credentials (username/password, bag or biometric data). The behavior identification of the user is needed to deduce information about the trust value while the capability identification is used to assign a priority value for user. The access control subclass defines a set of authorization (permit or deny), prohibition, recommendation and obligation which the decision is assigned according to the behavior, profile and context of the user. The policies are specified as a set of rules using SWRL form.

The User class: This entity aims to describe any person living with a disability due to any impairment, illness or simply aging.

The Profile subclass is used to provide fined grain/valuable data which help to distinguish one user from other by using both static profile (personal data, capabilities, hobbies) and dynamic profile which include (interests, preferences, opinions, moods). The capabilities subclass distinguishes cognitive, visual, hearing and motor impairments.

The Behavior subclass after having identified the specific profile and the capability of the person then we should check the behavior with respect to the background while classifying them into behavior class to recognize the right people.

The Context subclass describes the current state of the user in the environment that includes activity describing the current task, location characterizing the place of the user and time describing years, months, weeks, days, hours and minutes.

The Service class describes an entity solicited by an appropriate user capability and behavior.

The Device class aims to describe any item or piece of equipment used to maintain or improve the functional capabilities of a person with a disability. It focuses on assistive devices subclass including cognitive aids, hearing aids, visual aids, motor aids and resources subclass including audio, video and text.

The Environment class describes the intelligent and adaptable spaces that support the assistive services and devices for appropriate user profile capabilities like smart home, work space, learning and healthcare. This class specifies the indoor and outdoor subclasses.

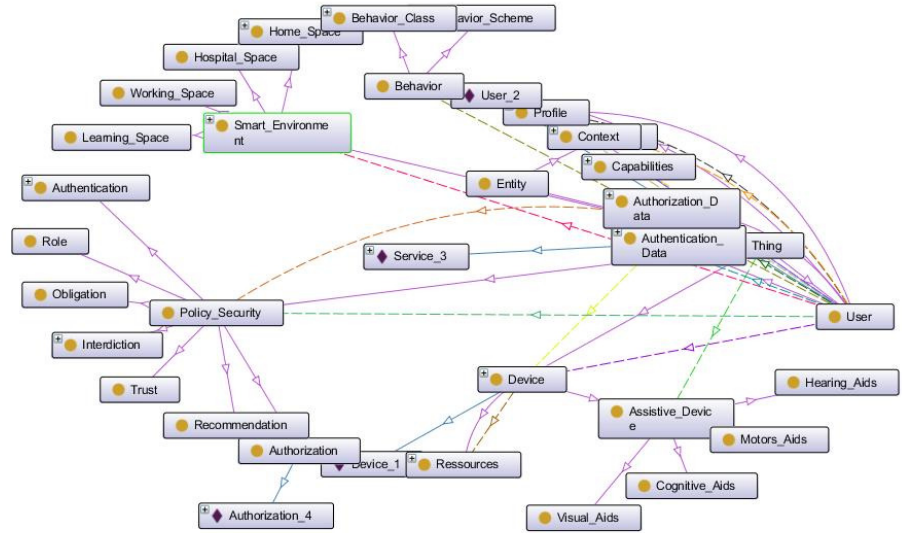

Figure 2.    Ontology model (UBC-ACM).

## 4.2. Reasoning

In order to infer the suitable authentication and access control decisions, we use a semantic reasoner that is able to infer the decision from a set of asserted observations about the dependant person located in smart environment. Therefore, these persons needs to access to smart services that take into account their impairments by using assistive devices. We specify semantic rules by the means of semantic web technologies. Usually, the rules respect the predicate logic and are specified in the form <if conditions then conclusion> to perform the reasoning. Semantic reasoner aims to check the consistency of our proposed ontology model and derive high and implicit knowledge about the situation, the profile, the capability, the authentication and the access control of the user.

We choose to implement the reasoning by using a complete open source OWL-DL reasoner as pellet, racer that conforms to our ontology modeling.

## 4.3. Rules specification

Our ontology is rich by using the contextual data, the capability and the behavior from users in such intelligent environment. This process is powered by the heterogeneous data captured from different sensors. We focus on ontology developed and rule base to perform the reasoning in order to deduce a new implicit knowledge which the rules are expressed as SWRL form. In this section, we organize the rules on contextual, profile capability, behavioral, authentication and access control rules specification. Contextual rules permit to provide a contextual knowledge about the user situation, time, location, device and environment. These data will be used in the remainder rules specification to deduce other high implicit knowledge. Profile capability rules, when a user asks for a particular service and device, we can deduce the user capability. For example, if the user uses an assistive device like hearing aids then we conclude that the user has hearing impairments and needs to input/output video resources. Behavioral rules provide to conclude other data about the profile, the authentication and the access control, if the user has a recognized behavior class. Authentication rules permit to deduce the decision according to the recognized user behavior and to confirm the decision, we use the traditional authentication means. Authorization rules need to validate the inferred data about the behavior, capability, context, service and device to assign the suitable decision to users.

### 4.4. Query specification

Once, the data are modeled and the rules are specified with rich semantics, therefore, we may query the stored data using a semantic query language like SPARQL.

In our approach, we use two kinds of queries. The first one to specify an authentication query that requires a user name, password, behavior model to pass the query to decision making level based on inference engine. The second to send an authorization query to inference engine with the specification of the situation, profile, capability and behavior.

## 5. PROPOSED ARCHITECTURE

In this section, we propose a new profile based Access Control Architecture which implements our model. The architecture is based on ontologies to ensure semi-automatic management and modeling. The architecture requires as input the profile, the behavior model and the current state of the user and it provides as output an authentication and authorization decision access. The architecture is built on the acquisition, the management context and the security layers as described in the figure below.

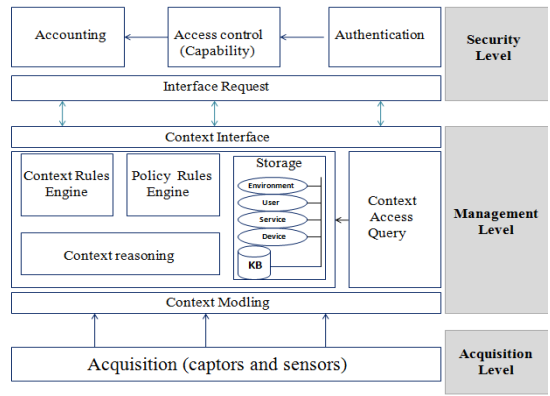

Figure 3. Proposed access control architecture

**Acquisition layer:** This layer captures the contextual data from hardware, software and combination of sensors on raw format that will be passed to the upper layer through an interface.

**Management layer:** This layer represents the architecture core that provides classifying, modeling, reasoning storage and querying process.

a) Context modeling: provides to make the entities in standard representation of context data using OWL language to enable interaction between different modules in our architecture.

b) Knowledge database: stores the data that permits to query, add, delete or modify context knowledge.

c) Policy, context rules engines: are expressed as a set of rules that will be used for inferring valid context from current situation.

d) Context reasoning engine: deduces new implicit knowledge.

e) Context access query: aims to invoke the database by specifying a SPARQL queries which are organized as authentication and access control queries.
The interaction between the different modules is described in this section:
The data are collected from physical and virtual sensors, the collected data will considered as data set test to predict the behavior class and generate the model class of data as recognized behavior class. The data will be represented and stored in the knowledge base. Authentication module sends an authentication query to the management layer. Look up the Inference Rules for defined rules which takes premices existed in the context and returns a conclusion; Context Reasoning Engine sends queries knowledge base. Response from Context Representation is given to Context Reasoning Engine and passed to context interface that will be transmitted to authentication module. Authorization access passed an AC request to Context Interface to collect the right decision. The log file will be updated over time.

**Security layer:** This layer is organized on three main processes: Authentication, Authorization and Accounting. It aims to ensure adaptive security policies in smart environment.

*Authentication:* to ensure an adaptive policy, the correct identification is required by checking the capability, the behavior then using the authentication means. The improved identification provides better personalization.

*Authorization:* once the user is identified correctly according to his profile and class of behavior, capability and the current contextual data then the suitable decision will be derived about the user.

*Accounting:* Tracking the user activities to detect the incorrect activity because the dependent users require a high assistance in their smart home to update the historic file.

The authentication policies allow identifying correctly the user depending on his capability and the recognized behavior. With the growing needs of personalization, the authentication process must be adaptive according to the capabilities, e.g., some people with physical impairments cannot use the username/password means.
The authentication policies are expressed in the form of rules which returns a "yes" or "no". When the person is recognized, we will have a trust value. Thus, an adaptive access control mechanism will be launched, this aim to specify the authorized persons to access to the services. The policies are expressed in the form of rules depending on various components, the users will be assigned to behavior and capability classes and the decision is validated according to the current context. Our adaptive policy will be able to reason on the behavior, the capability, the profile, the context in order to give a priority value and to enforce the security policy adaptation. Furthermore, the profile rules permit to deduce knowledge about the preferences, habits and hobbies. The access control rules aim to use all data to provide a "permit", "deny", "obligation" or "recommendation" decision. The management of these data is provided by an architecture that permits to gather data from sensors to enrich the ontology. The decision is acquired by specifying an authentication or access control queries to ask the ontology.

# 6. SCENARIOS

We choose to study three critical cases where a dependent person is facing a risky situation at home. Using our adaptive access control approach, we explain how we can strongly identify the user and derive a suitable access decision according to the user behavior and capability assuming a given service request and use of some specific assistive device.

Using our access control system, we go through three phases to derive the right decisions: modeling, reasoning and asking the ontology model.

| | |
|---|---|
| **Scenario 1:**<br><br>**Deaf Person** | **Situation:** pers on having hearing impairment.<br>                So, it needs to specific visual devices to perform their tasks as reading<br>              and daily activities….<br><br>          **Action:** Visual alert. |
| **Scenario 2:**<br>**Blind Person** | **Situation:** person having visual impairment.<br>            So, it needs to specific audible devices to perform their tasks as reading<br>            and daily activities easily ….<br>**Action:** Audible alert. |
| **Scenario 3:**<br>**Alzheimer Person** | **Situation:** It is a progressive disease which the person destroys cognitive<br>           abilities and it suffers about memory loss, abnormal behavior, a change in personality and an increase of anxiety.<br><br>**Action:** signaling the emergency situation and correcting the future activity. |

**Table .1.** Motivation proposed scenarios.

## 6.1 Modeling the security policy ontology

We propose an ontology model taking into account the three case studies: deaf person, blind person, Alzheimer person. These persons live in smart environment; they ask different services, use different devices according to the disability type with different context situation.

### 6.1.1 Authentication and authorization reasoning

We study the different use cases and showing the importance of an adaptive access control and the role of tracking the behavior, profile capability in such adaptive access control system. The reasoning is distinguished on two main classes: authentication and access control.

### 6.1.2 Authentication policies

Using some defined rules, we can ensure the identification of the right users and adapt the authentication means according to the dependency type. This process is ensured basing on capability type, behavior class and authentication means.
This rule contains activity, location and time as condition to deduce the recognized behavior.
   HasActivity(?u, xxx)^HasLocation (?u, yyyy) ^HasTime(?u, zzzz)
    →HasRecognizedbehavior(?u, class1)
This rule aims to deduce the capability according to the device used.

UsedDevice(?d, AssistedDevice)->HasCapability(?u, "yes")

This rule integrates the behavior and capability to deduce the appropriate mean that will be used in the authentication.

With rule 1, the user has not specific dependency then he can use the username and password mean.

Unlike rule 2, the user has a physical impairment then he cannot use the username password mean. For this, we propose to use tag mean to authentication.

**Rule 1:**
HasRecognizedbehavior(?u, class1)^HasCapability( ?u, "no")
->Authentication(username/password)
**Rule 2:**
HasRecognizedBehavior(?y, class2) ^ HasCapability( ?y, "physical")
->Authentication(tag-mean)

In this step, we check the identity of the user by using his name and password which the user has not capability.

Username(?u, kkkk)^Password(?u, hhhh)-> Authenticated(?u, yes).

### 6.1.3 Authorization policies

The access control is ensured according to the assignment of the users to behavior and capability groups then we check the valid time, location, device, service and environment to assign the "permit" or "deny" decision.

We use this rule to assign the user into groups according to their behavior and capabilities.

HasRecognizedbehavior(?u, class1)^ HasCapability( ?u, "hearing")
->BehaviorCapability(?u,Group1)^ HasRecognizedbehavior(?u, class2)^
HasCapability( ?u, "visual") ->BehaviorCapability(?u,Group2)

HasRecognizedbehavior(?u, "class2")^HasCapability( ?u, "cognitive")
->BehaviorCapability(?u,Group3)

**Case 1: Blind**
BehaviorCapability(?u,Group2)^AskedService(?Group2,)^UsedDevice(?Group,)^HasContext()
 ->has Access (?u, permit).
**Case 2: Deaf**
BehaviorCapability(?u,Group2)^ AskedService()^UsedDevice() ^HasContext()
 ->has Access (?u, permit).
**Case 3: Alzhiemer**
BehaviorCapability(?u,Group3)^ AskedService(?Group3, OpenDoor)^ HasContext(?time,
"00.00")
->has Access (?Group3, "Deny").

### 6.1.4 Quering The model

We send an authentication and authorization queries. Following this scenario, we have been shown the importance of taking into account the user behavior and capability when specifying the security policy and how adapting both authentication and access control security services from the needs of users.

## 7. CONCLUSION

Pervasive computing integrates more and more assistive technologies into every day space to give an easy and autonomy living to dependent people in smart environment. For this, the adaptation to the user needs becomes the key issue for providing the access to personalized services. In this paper, we have proposed an access control model based on behavior and capability which is based on the semantic web technologies (OWL language, SWRL for rule specification and SPARQL to query the ontology). Our security policy was developed with an ontology which expresses the concepts and rules security policy. The model supported by our proposed architecture is involving three design layers (acquisition, management and security). The major strength of our work is its ability to authenticate the users according to their behavior class. The access control to services is ensured according to the profile, capability and context of the user. The selected scenarios illustrate how the visual, the hearing and the cognitive capabilities can affect the design of security policy to ensure both authentication and access control security services. Motivated by these scenarios, we are planning to deploy our model and architecture in real living area of physically impaired people. To ensure the adaptability in such smart environment, we need to some personal data that will be shared. For this, the privacy and scalability must be taken into account in our future work.